# Optimizing TCP Loss Recovery Performance Over Mobile Data Networks

Ke Liu[1], Zhongbin Zha[1], Wenkai Wan[1], Vaneet Aggarwal[2], Binzhang Fu[1], Mingyu Chen[1]
[1]*State Key Laboratory of Computer Architecture, Institute of Computing Technology, Chinese Academy of Sciences*
[2]*Purdue University*

*Abstract* — Recent advances in high-speed mobile networks have revealed new bottlenecks in ubiquitous TCP protocol deployed in the Internet. In addition to differentiating non-congestive loss from congestive loss, our experiments revealed two significant performance bottlenecks during the loss recovery phase: flow control bottleneck and application stall, resulting in degradation in QoS performance. To tackle these two problems we firstly develop a novel opportunistic retransmission algorithm to eliminate the flow control bottleneck, which enables TCP sender to transmit new packets even if receiver's receiving window is exhausted. Secondly, application stall can be significantly alleviated by carefully monitoring and tuning the TCP sending buffer growth mechanism. We implemented and modularized the proposed algorithms in the Linux kernel thus they can plug-and-play with the existing TCP loss recovery algorithms easily. Using emulated experiments we showed that, compared to the existing TCP loss recovery algorithms, the proposed optimization algorithms improve the bandwidth efficiency by up to 133% and completely mitigate RTT spikes, i.e., over 50% RTT reduction, over the loss recovery phase.

## I. Introduction

Recent advances in high-speed mobile networks have revealed new bottlenecks in the ubiquitous TCP protocol deployed in the Internet. For example, Mascolo *et al.* [1] and Fu and Liew [2] first proposed to regulate the congestion window (CWnd) adaptively upon every loss event to differentiate the random packet loss in mobile networks from congestion loss, instead of reducing CWnd blindly. They showed that their methods are effective – unnecessary throughput degradations resulting from random losses can be mitigated, thus improving TCP's throughput performance.

More recently, adaptive loss recovery algorithms were invented and commonly adopted in numerous systems [3-6] and rate/congestion control algorithms [7-11], which have been used to replace the existing congestion control algorithms in conventional TCP. The key adaptive loss recovery algorithms can be divided into two categories: *queue length (or delay) adaptive loss recovery algorithm* and *bandwidth adaptive loss recovery algorithm*. The intuition behind them is to decouple the packet loss recovery from the congestion control hence their transmission rates or CWnds for both retransmissions as well as new data packets are only affected by either the estimated queueing length or bandwidth availability. Specifically, queue length adaptive loss recovery algorithm continuously adjusts its transmission rate or CWnd during the loss recovery phase so that the estimated queue length will oscillate around a target queue length. Bandwidth adaptive loss recovery algorithm tracks the bandwidth availability by estimating the DUPACK/ACK returning rate, and regulates the transmission rate/CWnd during the loss recovery phase adaptively. Compared to conventional TCP, these adaptive loss recovery algorithms can effectively alleviate the throughput degradation due to the congestion control bottleneck during the loss recovery phase.

Nonetheless, these works did not consider the *flow control bottleneck* in TCP's loss recovery phase – receiver's receiving buffer (AWnd) prevents new data packets from transmitting if AWnd is exhausted, resulting in bandwidth inefficiency. However, those new data packets will be transmitted in burst after loss recovery phase, resulting in the significant increase in RTT – *RTT spikes*. Recent measurements [3, 12] from production networks showed the prevalence of this problem during the TCP congestion avoidance phase, i.e., no loss occurred, but do not consider its performance impact during the loss recovery phase. Moreover, recent works that regulate the transmission rate using AWnd, such as DRWA [13] and RSFC [14], also suffer from the same problem.

To tackle this problem we develop a novel *opportunistic retransmission* algorithm that relaxes the AWnd constraint during TCP's loss recovery phase, hence allowing TCP sender to transmit new TCP segments as soon as the lost TCP segments (which have been detected so far) have been retransmitted irrespective of the AWnd size.

Interestingly, the throughput performance improves, but not as much as we expected. After investigating the trace data we discovered that TCP sender always runs out of packets – *application stall*. This always occurs during transmissions, especially during the loss recovery phase, thus TCP sender has no new data packets to transmit opportunistically. Our experiments showed that application stall occurs during over 61.1% of the loss recovery phase in practice. Dukkipati *et al.* [15] also demonstrated this problem but did not propose any solution. In this paper, we tackle this problem by refining the sending buffer growth mechanism to guarantee that TCP sending buffer always has data packets to transmit.

It is now increasingly common for mobile operators to deploy transparent proxies or servers inside the networks of mobile operators [3-6, 12-14, 16, 17]. The TCP flow will end-up terminating at the proxy/server. The mobile operators are very likely to deploy an optimized TCP stack/rate controller [7-11, 14, 18-20] at the proxies/servers for the last leg from the proxies/servers to the mobile device. Therefore, we could realize the proposed algorithms in those proxies/servers. Therefore, we modularized the proposed algorithms in the Linux kernel and thus they can plug-and-play with existing proxy/server's TCP implementations easily. This approach is practical because they do *not* require any modifications to the existing TCP implementation at the real client/server hosts, thus can be readily deployed into current mobile data networks.

We evaluate our proposed algorithms using emulated experiments, and show that, by using our proposed algorithms (a) the bandwidth over the loss recovery phase can be efficiently utilized and RTT spikes are mitigated completely; (b) the bandwidth efficiencies of all conventional TCP variants can be improved substantially and achieve near-optimal bandwidth utilization (over 89%) over mobile data networks.

The rest of the paper is organized as follows: Section II reviews the current TCP loss recovery algorithms. In Section III, we develop a system model to analyze their bandwidth utilization during the loss recovery phase. Section IV presents the opportunistic retransmission algorithm and the extended system model which is used to analyze the performance gain from opportunistic retransmission. Section V investigates the application stall problem and proposes a refined sending buffer growth mechanism. Section VI and VII present the implementation and experimental results to validate the performance gain resulting from these two optimization algorithms. Section VIII evaluates the impact of the assumption behind opportunistic retransmission on throughput performances. We summarize the work in Section IX.

## II. TCP LOSS RECOVERY REVISITED

We revisit the five existing loss recovery algorithms: (a) the standard TCP loss recovery algorithm as defined in RFC3517 [21]; (b) rate-halving [22]; (c) proportional rate reduction; (d) queue length adaptive rate reduction and (e) bandwidth adaptive rate reduction.

### A. Standard TCP Loss Recovery Algorithm

According to RFC3517, TCP enters the loss recovery phase either after a packet retransmission timeout or upon receiving *dupthresh* (typically three) number of duplicate ACKs (DUPACKs). In the timeout case, the CWnd size will be reduced to one and the TCP sender enters the slow start phase again. In this case, AWnd is likely larger than CWnd and thus unlikely to be the bottleneck. Therefore, we will not consider the timeout case in the rest of the paper.

In the second case, the TCP sender will enter the loss recovery phase and set both CWnd and slow start threshold (*ssthresh*) to equal to $0.5 \times FlightSize$ – where *FlightSize* is the amount of data that has been transmitted but not yet cumulatively acknowledged. Next, the TCP sender will retransmit the first unacknowledged TCP segment – known as fast retransmit.

If there are more than one lost segments reported in the SACK block [23], then the TCP sender will retransmit the remaining lost segments if the pipe is less than the congestion window, i.e., *pipe*<CWnd, where *pipe* is equal to *FlightSize* [24] (The amount of data that has been sent but not yet cumulatively acknowledged) minus the amount of out-of-sequence data reported as received by SACK blocks.

The loss recovery phase ends when the sender's highest sequence number at the time of entering the loss recovery phase is cumulatively acknowledged or when a timeout occurs.

After the TCP sender retransmits all lost segments reported by SACK blocks, it will transmit new TCP segments if both of the following conditions are met:

a) the amount of inflight data is less than the congestion window, i.e., *pipe*<CWnd;

b) the highest sequence number transmitted is less than the limit set by the receiver advertised window (AWnd).

Therefore the TCP sender may be prevented from sending new TCP segments if the AWnd becomes the bottleneck. As discussed in Section I, the AWnd is increasingly becoming the bottleneck.

### B. Rate-halving

Rate-Halving (RH) [22] is the default TCP fast recovery algorithm in Linux before kernel version 3.2, which still dominates the Linux kernel deployed in today's Internet servers. Thus, its loss recovery algorithm impacts a major portion of the Internet traffic.

The RH algorithm can be described in the following five steps:

a) The Linux TCP sender enters the loss recovery phase as soon as one out-of-order segment is guaranteed to be lost based on SACK [23] blocks and FACK [24] instead of receiving *dupthresh* DUPACK, e.g., *dupthresh*=3, described in RFC3517 [21]. As we only consider the loss recovery phase, the algorithm to determine whether to enter the loss recovery phase is out of the scope of this paper. Interested readers can refer to [25] for details.

b) Then it sets *ssthresh* to equal to $β \times CWnd$, where $β$ depends on the TCP variants used, e.g., $β$=0.7 for TCP CUBIC, $β$=0.5 for TCP Reno, $β$=0.8 for TCP Veno if the loss is determined to be non-congestive, otherwise, $β$=0.5. TCP Westwood is designed to avoid the blind *ssthresh* reduction due to random loss hence it sets *ssthresh* to equal to the product of the estimated bandwidth and the minimum RTT.

c) The CWnd size is set to *pipe*+1 if *pipe*+1<CWnd, where *pipe* is the amount of data outstanding in the network. This is designed to prevent transmission bursts (e.g., when CWnd–*pipe* is large) by limiting the sender to transmit at most one packet until a new ACK arrives.

d) If CWnd>*ssthresh*, TCP sender will reduce CWnd by one for every two ACKs received until CWnd=*ssthresh* – this is known as rate-halving and is designed to enable the sender to retransmit/transmit lost/new segments earlier and to space out the transmissions [22]. If CWnd<*ssthresh*, CWnd is limited by *pipe*+1 (as shown in step c) and cannot increase to *ssthresh*, which makes it possible to underutilize the bandwidth during the loss recovery phase as the target CWnd, *ssthresh*, is not satisfied.

e) The loss recovery phase ends when the sender's highest sequence number at the time of entering the loss recovery phase is cumulatively acknowledged.

When the TCP sender has retransmitted all lost segments reported by SACK blocks, it will transmit new TCP segments when the above two conditions, i.e., (a) and (b) in Section II.A, are met.

*C. Proportional Rate Reduction*

Proportional Rate Reduction (PRR) [15] has been introduced as the default TCP loss recovery algorithm in Linux kernel since version 3.2. The PRR differs from RH in two ways:

a) If CWnd>*ssthresh*, the TCP sender reduces CWnd by (1–*β*) for every received DUPACK/ACK until CWnd=*ssthresh* – this is known as proportional reduction. Like RH, PRR converges to the target CWnd, *ssthresh*, determined according to the used TCP variants, but at a higher convergence rate.

b) Similar to RH, PRR sets CWnd=*pipe*+1 if *pipe*<CWnd due to heavy packet losses [22]. Different from RH that remains CWnd at *pipe*+1 even if CWnd<*ssthresh*, PRR updates CWnd in the same manner as the TCP slow start phase, increasing CWnd by 1 for every received DUPACK/ACK until CWnd=*ssthresh*, which helps achieve the better bandwidth utilization during the loss recovery phase compared to RH.

Despite the above differences, the PRR sender still complies with the same two conditions, i.e., (a) and (b) in Section II.A, as in RH and standard TCP loss recovery algorithm, when transmitting new segments. Therefore PRR suffers from the same AWnd bottleneck as RH and standard TCP loss recovery algorithm.

*D. Queue Length Adaptive Loss Recovery Algorithm*

TCP Veno continuously estimates the queue length to differentiate random losses from congestion losses. Let $Q_i$ denote the queue length after receiving the $i^{th}$ ACK/DUPACK, and $Q_T$ denote the target queue length. Upon a loss event, TCP Veno treats that loss as a random loss and sets *ssthresh* to 0.8CWnd if $Q_i \leq Q_T$, because it considers the network is not congestive. Otherwise (i.e., $Q_i > Q_T$), TCP Veno treats that loss as a congestion loss thus sets *ssthresh* to 0.5CWnd. However, its CWnd is still governed by the loss recovery algorithms used, e.g., RH and PRR, during the loss recovery phase.

To alleviate this congestion control bottleneck during the loss recovery phase, Leong *et al.* [7], Liu and Lee [8], and Ren and Lin [4] developed a queue length congestion control algorithm that estimates the backlog at the bottleneck link and limits it to a target queue length by adjusting CWnd or the transmission rate accordingly during the whole transmission *including the loss recovery phase*. Hence, its transmission rate or CWnd is only affected by the queue length thus packet losses are decoupled from CWnd or rate controls.

As we cannot find the source code of the current queue length adaptive loss recovery implementation, we develop a Queue length Adaptive Rate Reduction algorithm (QARR) in Linux kernel based on the similar idea. The QARR is different from RH and PRR in the following three ways:

a) QARR does not employ *ssthresh*, thus its CWnd will not converge to *ssthresh* but is only affected by the estimated queue length.

b) After entering the loss recovery phase, QARR adopts the Vegas/Veno method to estimate the queue length for every received ACK/DUPACK. Let $RTT_i$ denote the RTT measured after receiving the $i^{th}$ ACK/DUPACK using the TCP timestamp option appended in the received ACK/DUPACK's TCP header; $pipe_i$ denote the number of inflight packets after receiving the $i^{th}$ ACK and *baseRTT* denote the minimum RTT during the transmission. Hence $Q_i$ can be estimated as follows:
$$Q_i = pipe_i(RTT_i - baseRTT)/RTT_i \quad (1)$$

c) Then QARR regulates the queue length around a target queue length by adjusting CWnd. Let $W_i$ denote the CWnd after receiving the $i^{th}$ ACK/DUPACK, then $W_i$ is determined as follows:
$$W_i = W_{i-1} - \max(Q_i - Q_T, 0) \quad (2)$$

QARR also does not consider the AWnd bottleneck, i.e., condition (a) and (b) in Section II.A, as in RH and PRR when transmitting new segments.

*E. Bandwidth Adaptive Loss Recovery Algorithm*

TCP Westwood continuously estimates the bandwidth to differentiate mobile networks' random packet losses from congestion losses. Let $B_i$ denote the estimated receiving bandwidth after receiving the $i^{th}$ ACK/DUPACK. TCP Westwood sets its *ssthresh* to $B_i \times RTT_{min}$, i.e., the effective bandwidth delay product (BDP), after the loss recovery phase. Similar to TCP Veno, its CWnd is also governed by RH/PRR during the loss recovery phase. Therefore, recent works [3, 5, 6, 9-11] proposed to decouple the packet loss recovery from the congestion/rate control, so CWnd/the transmission rate during the loss recovery phase, in this case, is only governed by the estimated bandwidth.

We also develop a Bandwidth Adaptive Rate Reduction algorithm (BARR) in Linux kernel based on the similar idea. BARR is different from RH and PRR in the following ways:

a) BARR does not employ *ssthresh*, thus CWnd will not converge to *ssthresh* but is only affected by the estimated bandwidth.

b) BARR continuously performs an estimate of the available bandwidth by measuring the average rate of returning ACKs. Specifically, let $t_i$ be the arrival time of ACK $i$ with acknowledged sequence number $ack_i$. Then for a positive integer $M$, $B_i$ is computed from
$$B_i = \frac{ack_{i+M} - ack_i}{t_{i+M} - t_i} \quad (3)$$

where the numerator is the amount of data received by the receiver during the time interval $(t_i, t_{i+M})$. The parameter $M$ controls the duration of the estimation interval (in number of ACKs) and can be adjusted to a tradeoff between accuracy and timeliness of rate estimations.

c) BARR sets $W_i$ at the $i^{th}$ received ACK according to $B_i$ in (3). Thus, $W_i$ is computed from
$$W_i = B_i \times RTT_i \quad (4)$$

Similarly, BARRs in previous works also suffer from the AWnd bottleneck i.e., condition (a) and (b) in Section II.A, as in RH and PRR.

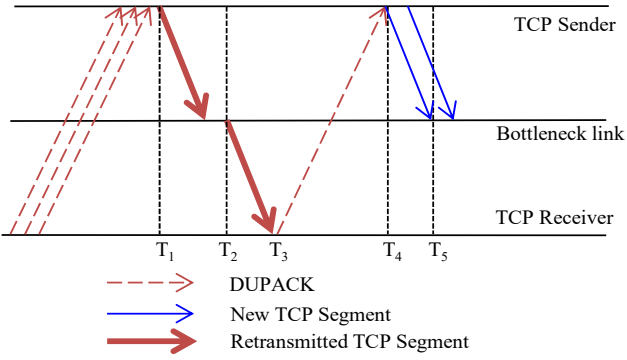

Fig. 1 – System model in loss recovery period.

## III. SYSTEM MODEL

In this section, we develop a system model to investigate the efficiency of the five loss recovery algorithms reviewed in Section II. Specifically, we are interested in the efficiency at which these loss recovery algorithms utilize the available link bandwidth during the loss recovery phase, and quantifies the impact of various system parameters, most notably the AWnd size, on the link utilization efficiency.

We consider a system where a TCP sender (e.g., an Internet server) transmits data over a mobile network to a TCP receiver (e.g., mobile handset), as shown in Fig. 1.

We assume the mobile network is the bottleneck link and that link maintains a queue for packets to be transmitted to the receiver. To efficiently utilize the bottleneck link the queue must not become empty. Hence our goal is to quantify the amount of time the queue is empty during the loss recovery phase – link idle time.

In the following we consider a single TCP flow where:
 a) it lasts longer than a loss recovery phase;
 b) all TCP segments are of the same size;
 c) the bottleneck link bandwidth remains constant during the loss recovery phase, let $C$ denote the bandwidth during the loss recovery phase (in packets per second);
 d) the RTT in the system model includes propagation delay, transmission delay and queuing delay, we assume that propagation delay remains constant during the loss recovery phase and let $U$ denote the propagation delay;
 e) TCP flow is not source-limited, i.e., the sender sends as much data as CWnd and AWnd allow;
 f) the receiver's receiving window size is not smaller than the network's bandwidth propagation delay product. Let $AW$ denote the AWnd size, thus $AW \geq CU$;

Furthermore, at the time the TCP flow enters the loss recovery phase, we assume that:
 g) the AWnd size remains constant during the loss recovery phase;
 h) the growth of the CWnd size is limited by $AW$, let $CW$ denote the CWnd size;

After entering the loss recovery phase, the sender retransmits the first lost packet reported in the SACK blocks. Further DUPACKs from the receiver will trigger the retransmissions of the remaining lost packets. Note that whenever a retransmitted packet is successfully recovered, it will enable the sender to transmit a new TCP segment once all lost packets have been retransmitted.

In the following analysis, we focus on the loss recovery period from the time instant the first retransmitted packet is sent by the bottleneck link ($T_2$ in Fig. 1), to the time instant the first new packet is sent by the bottleneck link (the time after or at $T_5$ in Fig. 1). In particular, we first derive the *minimum link idle time* – the minimum period the link is not transmitting packet, denoted by $I_{min}$, where the minimum link idle time is achieved when the lost packets are consecutive – *best case scenario* (because all the lost packets in this case are decoded before $T_2$ thus can be retransmitted back to back). Then we use it to compute the *maximum bandwidth utilization*, denoted by $\eta$ ($0 \leq \eta \leq 1$), for each of the five loss recovery algorithms, which measures the proportion of available bandwidth utilized for packet transmission under the best-case scenario.

### A. Standard TCP Loss Recovery Algorithm

The minimum link idle time for the standard TCP loss recovery algorithm [22] is stated in Theorem 1 below.

*Theorem 1: The minimum link idle time for the standard TCP loss recovery algorithm in the loss recovery period of $n$ lost packets is given by:*

$$I_{min} = \begin{cases} \max\{U-(n-1)/C, 0\} & n < \lceil 0.5AW \rceil \\ \max\{U-(\lceil 0.5AW \rceil -1)/C, 0\} & n \geq \lceil 0.5AW \rceil \end{cases} \quad (5)$$

*Proof:*

Let $n$ denote the number of the lost TCP segments in a loss recovery period. Consider the period from $T_2$ to $T_5$ as depicted in Fig. 1. Ignoring processing delay the duration from $T_2$ to $T_5$ is equal to one round-trip propagation delay, i.e., $U$ seconds. As the sender must retransmit all lost packets before transmitting a new TCP segment at $T_4$, it can at most send out $n-1$ retransmitted packets during this interval, subject to the CW. Note that upon entering the loss recovery phase TCP will set $CW = \lceil 0.5AW \rceil$. Next, we consider two cases:

Case 1: $n < CW = \lceil 0.5AW \rceil$: In this case, the sender can send out at most $n-1$ retransmitted packets during the interval $[T_2, T_5]$, with a total transmission time of $(n-1)/C$. Thus the minimum link idle time can then be computed from

$$I_{min} = \max\{U-(n-1)/C, 0\} \quad (6)$$

Case 2: $n \geq CW = \lceil 0.5AW \rceil$:

In this case, the sender can retransmit at most $\lceil 0.5AW \rceil$ lost packets subject to the CWnd constraint. Therefore the corresponding minimum link idle time is given by

$$I_{min} = \max\{U-(\lceil 0.5AW \rceil -1)/C, 0\} \quad (7)$$

In the best case scenario there is no waiting time for retransmitting the rest of lost packets and the 1st new packet after $T_5$, so (7) gives the overall minimum link idle time and the result follows. ∎

With the minimum link idle time, we can then compute the maximum bandwidth utilization from

$$\eta = n/(CI_{min}+n) \quad (8)$$

where the numerator is the amount of data transmitted (i.e., $n-1$ retransmitted packets and 1 new packet) and the denominator is the total amount of data that could be transmitted (i.e., amount transmitted plus the amount which would have been transmitted during the link idle time) during the loss recovery period.

### B. Rate-halving

The minimum link idle time for rate-halving loss recovery algorithm is stated in Theorem 1 below.

*Theorem 2: The minimum link idle time for RH's loss recovery algorithm is given by:*

$$I_{\min} = \begin{cases} \max\{U-(n-1)/C, 0\}, & n \leq \lfloor 0.5AW \rfloor \\ \max\{\lfloor n/(AW-n) \rfloor (U-(AW-n-1)/C), 0\}, & n > \lfloor 0.5AW \rfloor \end{cases} \quad (9)$$

*Proof: The proof is detailed in Appendix I.A.*

Similarly, the maximum bandwidth utilization can be computed from (8).

### C. Proportional Rate Reduction

The minimum link idle time for PRR loss recovery algorithm is stated in Theorem 2 below.

*Theorem 3: The minimum link idle time for PRR's loss recovery algorithm is given by:*

$$I_{\min} = \begin{cases} \max\{U-(n-1)/C, 0\}, & n \leq 2(AW-n) \\ \max\left\{ \sum_{i=1}^{\min k \atop s.t. \sum_{j=1}^{k}\min(\lceil \beta AW \rceil,(AW-n)2^j) \geq n} \left(U-\left(\min(\lceil \beta AW \rceil,(AW-n)2^i)-1\right)/C\right), 0 \right\}, & n > 2(AW-n) \end{cases} \quad (10)$$

*Proof: The proof is detailed in Appendix I.B.*

Similarly, the maximum bandwidth utilization can be computed from (8).

### D. Queue Length Rate Adaptive Reduction

The minimum link idle time for QARR loss recovery algorithm is stated in Theorem 3 below.

*Theorem 4: The minimum link idle time for QARR's loss recovery algorithm is given by:*

$$I_{\min} = \max\{U-(n-1)/C, 0\} \quad (11)$$

*Proof: The proof is detailed in Appendix I.C.*

Similarly, the maximum bandwidth utilization can be computed from (8).

### E. Bandwidth Adaptive Rate Reduction

Similar to QARR, BARR loss recovery algorithm has the same minimum link idle time, which is stated in Theorem 4 below.

*Theorem 5: The minimum link idle time for BARR's loss recovery algorithm is given by:*

$$I_{\min} = \max\{U-(n-1)/C, 0\} \quad (12)$$

*Proof: The proof is detailed in Appendix I.D.*

Similarly, the maximum bandwidth utilization can be computed from (8).

### F. Discussions

The previous analysis reveals two properties of the five loss recovery algorithms. First, in a network with high bandwidth (i.e., $C$) and long delay (i.e., $U$), the link will be more likely become idle during the loss recovery phase (c.f. (5), (9), (10), (11) and (12)).

Current 3G/HSPA+ networks have a typical bandwidth of 20Mbps and a delay of 100ms. Thus within one propagation delay (i.e., $U$) there is sufficient bandwidth to transmit 20Mbps×100ms/8bits=250KB data. With a typical packet size of around 1.5KB it is clear that the link will likely become idle during loss recovery unless the loss event comprises a burst of over 167 lost packets.

Moreover, the deficiency increases with higher bandwidth and thus will become even more significant in the emerging LTE/4G and 5G networks. In the next section, we propose a novel opportunistic retransmission algorithm to tackle this challenge.

## IV. OPPORTUNISTIC RETRANSMISSION

Liu and Lee [3, 26] proposed *opportunistic transmission* to tackle the flow control bottleneck in mobile networks with a large bandwidth-delay product (BDP). In particular, in networks with BDP larger than the receiver advertised window (AWnd). The motivation is that TCP's flow control algorithm was designed to prevent fast senders from overflowing slow receivers. However, rapid advances in processors have equipped modern day receivers such as PCs and smartphones with very high processing power, and thus the ability to process incoming packets at very high data rates. As a result, the AWnd size reported by the receiver is often kept at the maximum value, as few packets require extensive buffering to wait for processing.

Opportunistic transmission exploits the receiver's processing power by allowing the sender to transmit packets *beyond* the maximum sequence number allowed by AWnd. In practice, by the time the *out-of-AWnd* packets arrive at the receiver, the previously received packets would have been processed already, thus allowing the receiver to receive them without buffer overflow. But opportunistic transmission was only designed for TCP's normal phase of operation, i.e., when there is no packet loss occurred. Therefore, we extend the idea to TCP's loss recovery phase – *opportunistic retransmission*.

Specifically, as shown in Fig. 1, the TCP sender has to wait for a cumulative ACK with an increase in the highest sequence number acknowledged after retransmitting the lost packets due to AWnd constraint. If the lost packets are recovered successfully, the application at the receiver can process the packets quickly and free up the receiving window/buffer immediately for new packet arrivals if the receiver has sufficient processing power, e.g., at time $T_3$ in Fig. 1. However, the sender has to wait for the ACK acknowledging the first retransmitted packet to return before it can begin transmitting a new packet at time $T_4$ in Fig. 1, resulting in the link idle time analyzed in Section III.

Table. 1 – Comparison of bandwidth utilization during the loss recovery period versus different packet loss burst sizes.

| Loss recovery algorithm | | | Burst size (packets) | | |
|---|---|---|---|---|---|
| | | | 10 | 50 | 100 |
| RH | Original | Numerical | 0.054 | 0.294 | 0.474 |
| | | Experimental | 0.057 | 0.286 | 0.513 |
| | w/ OR | Numerical | 0.75 | 0.75 | 0.474 |
| | | Experimental | 0.485 | 0.737 | 0.437 |
| PRR | Original | Numerical | 0.054 | 0.294 | 0.594 |
| | | Experimental | 0.057 | 0.286 | 0.571 |
| | w/ OR | Numerical | 0.75 | 0.75 | 0.75 |
| | | Experimental | 0.383 | 0.703 | 0.731 |
| QARR | Original | Numerical | 0.054 | 0.294 | 0.594 |
| | | Experimental | 0.057 | 0.286 | 0.513 |
| | w/ OR | Numerical | 1 | 1 | 1 |
| | | Experimental | 0.246 | 0.543 | 0.983 |
| BARR | Original | Numerical | 0.054 | 0.294 | 0.594 |
| | | Experimental | 0.057 | 0.481 | 0.738 |
| | `w/ OR | Numerical | 1 | 1 | 1 |
| | | Experimental | 0.517 | 0.771 | 0.990 |

Table. 2 – Comparison of bandwidth utilization during the loss recovery period versus different propagation delays.

| Loss recovery algorithm | | | Propagation delay (ms) | | |
|---|---|---|---|---|---|
| | | | 50 | 150 | 200 |
| RH | Original | Numerical | 0.108 | 0.036 | 0.027 |
| | | Experimental | 0.114 | 0.038 | 0.028 |
| | w/ OR | Numerical | 0.75 | 0.75 | 0.75 |
| | | Experimental | 0.479 | 0.225 | 0.720 |
| PRR | Original | Numerical | 0.108 | 0.036 | 0.027 |
| | | Experimental | 0.114 | 0.038 | 0.028 |
| | w/ OR | Numerical | 0.75 | 0.75 | 0.75 |
| | | Experimental | 0.74 | 0.57 | 0.726 |
| QARR | Original | Numerical | 0.108 | 0.036 | 0.027 |
| | | Experimental | 0.114 | 0.038 | 0.028 |
| | w/ OR | Numerical | 1 | 1 | 1 |
| | | Experimental | 0.411 | 0.256 | 0.340 |
| BARR | Original | Numerical | 0.108 | 0.036 | 0.027 |
| | | Experimental | 0.113 | 0.035 | 0.033 |
| | w/ OR | Numerical | 1 | 1 | 1 |
| | | Experimental | 0.676 | 0.555 | 0.440 |

To tackle this problem, we propose to relax the AWnd constraint during TCP loss recovery phase by allowing transmitting new packets beyond the AWnd constraint, e.g., transmit new packets during $[T_1, T_4]$ in Fig. 1, irrespective of the AWnd. However, this must performed judiciously as the receiver is temporarily unable to clear its buffer (by passing data to the application) until the head-of-line lost packet is successfully retransmitted. Thus, opportunistic retransmission operates according to the following three ways:

a) For every received DUPACK the TCP sender decodes the SACK blocks carried inside DUPACK to determine (i) the number of gaps (i.e., lost packets) at the receiver's receiving buffer, denoted by $n_1$; and (ii) the number of out-of-order packets received, denoted by $n_2$.

b) Then the TCP sender first retransmits the lost packets. The receiver will advance the AWnd by $n_1+n_2$ packets if they are successfully received. Therefore the TCP sender then transmits up to $n_1+n_2$ new packets.

c) But the number of transmitted packets is also subject to the congestion control constraint: $pipe \leq CWnd$.

In the following we apply this opportunistic retransmission algorithm to the five loss recovery algorithms and derive their minimum link idle time and maximum bandwidth utilization.

### A. Performance Analysis

Opportunistic retransmission enables the five loss recovery algorithms to remove constraint (b) in Section II. The following five theorems show the minimum link idle time when using opportunistic retransmission.

Theorem 6 below states the minimum link idle time for standard loss recovery with opportunistic retransmission.

*Theorem 6: The minimum link idle time for standard loss recovery with opportunistic retransmission is given by:*

$$I_{\min} = \max\{U - (\lceil 0.5AW \rceil - 1)/C, 0\} \quad (13)$$

*Proof: The proof is detailed in Appendix I.E.*

Theorem 7 below states the minimum link idle time for rate-halving loss recovery algorithm with opportunistic retransmission.

*Theorem 7: The minimum link idle time for rate-halving loss recovery with opportunistic retransmission is given by:*

$$I_{\min} = \begin{cases} \max\{U - (\lceil \beta AW \rceil - 1)/C, 0\}, & n \leq AW - \lceil \beta AW \rceil \\ \max\{U - (AW - n - 1)/C, 0\}, & AW - \lceil \beta AW \rceil < n \leq AW - n \\ \max\{\lfloor n/(AW - n) \rfloor (U - (AW - n - 1)/C), 0\}, & n > AW - n \end{cases}$$

(14)

*Proof: The proof is detailed in Appendix I.F.*

Theorem 8 below states the minimum link idle time for proportional rate reduction loss recovery algorithm with opportunistic retransmission.

*Theorem 8: The minimum link idle time for proportional rate reduction loss recovery with opportunistic retransmission is given by:*

$$I_{\min} = \begin{cases} \max\{U - (\lceil \beta AW \rceil - 1)/C, 0\}, & n \leq 2(AW - n) \\ \max\left\{ \sum_{i=1}^{\min k \atop s.t. \sum_{j=1}^{k} \min(\lceil \beta AW \rceil, (AW-n)2^j) \geq n} (U - (\min(\lceil \beta AW \rceil, (AW-n)2^i) - 1)/C), 0 \right\}, & n > 2(AW - n) \end{cases}$$

(15)

*Proof: The proof is detailed in Appendix I.G.*

Theorem 9 below states the minimum link idle time for both queue length adaptive rate reduction and bandwidth adaptive rate reduction loss recovery algorithms with opportunistic retransmission.

*Theorem 9: Both queue length adaptive rate reduction loss recovery and bandwidth adaptive rate reduction loss recovery with opportunistic retransmission will not result in any link idle time, thus $I_{min}=0$.*

*Proof: The proof is detailed in Appendix I.H.*

Additionally, opportunistic retransmission can work based on the assumption that the mobile device has sufficient processing power, e.g., Apple A7 has a 64bit 1.3-1.4GHz dual-core CPU [27]. To experimentally validate this assumption we use cpulimit [28] to restrict the CPU utilization of a TCP process/application (e.g., *iperf*). We found that opportunistic retransmission still can work even if CPU utilization of the TCP process is limited to 10%.

*B. Performance Validation*

Using the system model we can evaluate the bandwidth utilization for those five loss recovery algorithms, and investigate the potential improvements achieved by using opportunistic retransmission. First, we compute the numerical results using the system model. Then, we verify the system model by experiments. Specifically, we adopt the typical system parameters of 3G/HSPA+ networks in the experimental setup. The system parameters are summarized in Table 3 and the experimental setup is shown in Fig. 2 (c.f., Section VII.A). We capture the TCP trace at the sender to measure the actual link idle time thus derive the bandwidth utilization during the loss recovery period. This enables us to verify the system model by comparing the experimental results to the numerical ones. As the standard TCP loss recovery was not implemented by the major operating systems we did not validate it by experiments.

To model loss events we test different numbers of consecutive packet losses, ranging from 10 to 100, to simulate the loss events under different network setting and radio signal conditions. Table 1 evaluates the impact of the burst size of packet loss events, ranging from 10 to 100 packets, on bandwidth utilization achieved over the loss recovery phase. We first observe that the numerical results computed from the system model are quite consistent with the experimental results for original TCP loss recovery algorithms under different loss rates, and the bandwidth utilization increase for larger burst size – a direct result of longer link utilization time to retransmit the larger number of lost packets, e.g., the bandwidth utilizations of QARR increase from 0.057 to over 0.51 with the increase in burst size of lost packets. However, the numerical results from the system model do not match the experimental results for TCP loss recovery algorithms with opportunistic retransmission under small burst size, e.g., 0.246 versus 1 for QARR under 10 burst size of lost packets. The similar observation can be found in Table 2 that compares the average bandwidth utilization over the loss recovery phase under different propagation delays ranging from 50ms to 200ms, when 10 consecutive packet loss event occurs, the numerical results are consistent with the experimental results for the original loss recovery algorithms under all the propagation delays, while they are inconsistent after applying opportunistic retransmission.

Interestingly, the throughputs of the existing loss recovery algorithms are improved by opportunistic retransmission but not as much as shown in our system model. The gap between the numerical results and the experimental results is larger with the smaller burst size of lost packets. After investigating the data trace we found that *application stall* (where TCP sender runs out of packets) always occurs during loss recovery phase. Hence, TCP sender has no new packet to transmit opportunistically, resulting in the degradation on the bandwidth utilization. Our experiments show that application stall occurs during over 61.1% of the loss recovery phase. In the next section, we refine the sender's *sndbuf* growth mechanism to prevent application stall from occurring.

V. APPLICATION STALL

Application stall can occur due to (a) application has no new data to send, and (b) the inefficiency of the current sending buffer growth mechanism. For (a) the sender does not always have data to send, which violates the assumption in Section III that TCP flow is not source-limited, hence we only consider (b). We first define *application stall* as follows:

*Definition 1: In a TCP data flow, the TCP sender cannot timely move the application data from the application layer to the transport layer, resulting in the transport layer not having sufficient data to transmit.*

In the following sections, we first investigate the TCP sender's sending process in current Linux TCP/IP stack and found that the application stall is mainly caused by the memory management of the sending buffer, i.e., *sndbuf*. Therefore, we first introduce the use of the *sndbuf* and its allocation strategy, and why it could result in the application stall problem. Then, to address this problem, we propose a refined sending buffer growth mechanism.

*A. Linux TCP sndbuf Memory Management*

When the user-space application requests to send data, the function *tcp_sendmsg()* is called in the kernel-space, which is responsible for passing data from the application layer to the transport layer only if the transport layer has enough memory to store that data. Since TCP is reliable, all sent data must be stored in the *sndbuf* before being sequentially acknowledged. Therefore, *tcp_sendmsg()* first checks whether there is enough memory to store that data by calling *tcp_memory_free()*. If no enough available memory, it adds *SOCK_NOSPACE* flag to the socket and goes into the *sleep* state until memory is available. If there is enough memory, it allocates a socket buffer (SKB, SocKet Buffer) for that data and passes that data to the transport layer, i.e., TCP, where SKB is the smallest unit in managing data. When a new TCP packet (e.g., ACK) is received by the TCP sender, *tcp_clean_rtx_queue()* is called to clear the data in the *sndbuf*. If the *SOCK_NOSPACE* flag was previously set, function *tcp_write_space()* is called to wake up the process in the *sleep* state and continue to pass the application data to the *sndbuf*. However, the process is awakened up only if the free space of the *sndbuf* is less or equal to the 1/3 of the total space. This setting can be the primary factor resulting in application stall, which will be explained in the next sections.

*B. Current Sending Buffer Growth Mechanism*

All the transmitted/transmitting but not acknowledged segments will be stored in the *sndbuf*, and will not be freed unless they are acknowledged sequentially. In Linux TCP, the size of *sndbuf* is not fixed but dynamically configured. The growth mechanism of *sndbuf* can be summarized in the following three ways:

a) The size of *sndbuf* is affected by the parameters in file *tcp_wmem* in the directory */proc/sys/net/ipv4*. It contains three values: the minimum *sndbuf* size, *snd_wmem_min*, the default *sndbuf* size, *snd_wmem_default*, and the maximum *sndbuf* size, *snd_wmem_max*. The *sndbuf* size is initialized to be *snd_wmem_default*, e.g., 16KB, in Centos 5.5.

b) *sndbuf* increases multiplicatively according to the CWnd size and the increasing factor is 2, which makes *sndbuf*=2*CW* at all times. *sndbuf* increases in this manner until CWnd stops increasing, being limited by AWnd, thus at this steady state, *sndbuf* size is twice of the current CWnd.

c) The segments occupied in the *sndbuf* will be freed if they are acknowledged sequentially, resulting the free space (*free_space*) in the *sndbuf*. Once *free_sapce*>*sndbuf*/3, *sndbuf* will move forward for *free_space* to fetch new data segments from application to fill it up.

However, this *sndbuf* growth mechanism cannot completely prevent application stall from occurring, especially during the loss recovery phase, with opportunistic retransmission. When *free_sapce*≤*sndbuf*/3, the application cannot send any data, i.e., application stall occurs, which could lead to two problems: (a) no packet is transmitted when the free space is released, resulting in the bandwidth inefficiency; (b) when the new data packets are passed to *sndbuf*, they will be sent in burst, which affects the accuracy of the queueing delay/RTT estimations in rate controls [4, 7, 8].

Hence we have the following theorem.

*Theorem 10*: Let $b_{free}$ denote the *free_space* size of the *sndbuf*. Assuming the sndbuf is in the steady state and AWnd is the bottleneck, with opportunistic retransmission, application stall will not occur during TCP loss recovery phase if $b_{free} \leq (1-\beta)CW$ or $b_{free} > 2CW/3$. Therefore, if $(1-\beta)CW < b_{free} \leq 2CW/3$, application stall occurs during the TCP loss recovery phase.

*Proof*:

Let $b_{snd}$ denote the *sndbuf* size. Since the *sndbuf* is in the steady state, we have $b_{snd}=2CW$. Let $b_{occ}$ denote the size of the *sndbuf* being occupied by the data, thus $b_{occ}=b_{snd}-b_{free}$. During the loss recovery phase TCP sender can at most transmit $\beta CW$ new segments using opportunistic retransmission. To satisfy it we need

$$b_{occ}=b_{snd}-b_{free} \geq (1+\beta)CW \quad (16)$$

thus $b_{free} \leq (1-\beta)CW$.

As shown in (c) – if $b_{free}>b_{snd}/3$, *sndbuf* will advance *free_space* to fetch new data to fill it up, thus $b_{occ}=b_{snd}$ and $b_{free}=0$ after fetching the data. As $b_{occ}=b_{snd} \geq (1+\beta)CW$, TCP sender also can at most transmit $\beta CW$ new segments if

$$b_{free}>b_{snd}/3=2CW/3 \quad (17)$$

Therefore, application stall will not occur if either (16) or (17) is satisfied, and application stall can occur if $(1-\beta)CW < b_{free} \leq 2CW/3$. ∎

The reduction factor $\beta$ affects the period of the application stall. For TCP CUBIC and PRR combination, $\beta$=0.7, application stall can occur if $0.3CW < b_{free} \leq 2CW/3$, thus $(2/3-0.3)/(2/3)$=55% of the time application stall can occur theoretically, while 61.6% of the time application stall occurs from our measurements. Similarly, for TCP CUBIC and QARR/BARR combinations, $\beta$=1, application stall can occur if $0 < b_{free} \leq 2CW/3$, thus 100% of the time the application stall can occur during the loss recovery phase, which is also validated in Section IV.B.

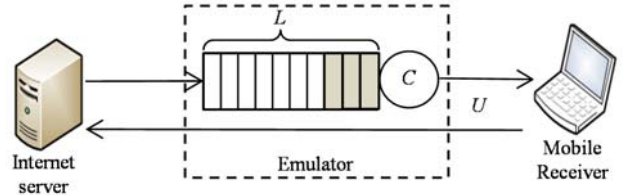

Fig. 2 – Experiments testbed setup

Table. 3 – System parameters and experimental setups.

| Parameters | Values | Description |
|---|---|---|
| $U$ | 50ms,100ms,150ms,200ms | Propagation delay |
| $L$ | 4.3MB [3] | Link buffer size |
| $P$ | 200s | Experiment period |
| $C$ | 20Mbps/100Mbps | Bandwidth for HSPA+/LTE |
| $S$ | 1515B | Packet size |
| $AW$ | $C \times U$ | Receiver's receiving window |
| # of runs | 10 | Number of runs |
| $\rho$ | 0.01%, 0.05%, 0.1% | Packet loss rate |

Table. 4 – Comparison of bandwidth utilization during the loss recovery period versus different loss.

| Loss recovery algorithm | | Burst size (packets) | | |
|---|---|---|---|---|
| | | 10 | 50 | 100 |
| RH | w/ the proposed optimization algorithms Numerical | 0.75 | 0.75 | 0.47 |
| | Experimental | 0.74 | 0.74 | 0.51 |
| PRR | Numerical | 0.75 | 0.75 | 0.75 |
| | Experimental | 0.74 | 0.74 | 0.74 |
| QARR | Numerical | 1 | 1 | 1 |
| | Experimental | 0.96 | 0.98 | 0.95 |
| BARR | Numerical | 1 | 1 | 1 |
| | Experimental | 0.97 | 0.98 | 0.96 |

Table. 5 – Comparison of bandwidth utilization during the loss recovery period versus different propagation delays.

| Loss recovery algorithm | | Propagation delay (ms) | | |
|---|---|---|---|---|
| | | 50 | 150 | 200 |
| RH | w/ the proposed optimization algorithms Numerical | 0.75 | 0.75 | 0.75 |
| | Experimental | 0.74 | 0.73 | 0.73 |
| PRR | Numerical | 0.75 | 0.75 | 0.75 |
| | Experimental | 0.74 | 0.73 | 0.73 |
| QARR | Numerical | 1 | 1 | 1 |
| | Experimental | 0.95 | 0.96 | 0.98 |
| BARR | Numerical | 1 | 1 | 1 |
| | Experimental | 0.99 | 0.95 | 0.98 |

### C. Refined Sending Buffer Growth Mechanism

Intuitively, we can remove the constraint in (c) in Section V.B, that is, once *free_space* is available *sndbuf* will advance and fetch new data from application to fill it up. However, this method will lead to significant extra scheduling delay hence we solve it in another way.

We modify the *sndbuf* increasing factor from 2 to 3, i.e., *sndbuf*=3× current CWnd size, and we can show that 3 is the minimum increasing factor that guarantees application stall would not occur during loss recovery phase, thus we have the following theorem.

*Theorem 11*: Application stall will not occur if *sndbuf* increases multiplicatively and the increasing factor is at least 3, thus $b_{snd}$=3*CW*.

*Proof*: if without application stall, we have

$$b_{snd} \geq \max(b_{free}+b_{occ}) \quad (18)$$

So we have

$$b_{snd} \geq \max(b_{free})+\max(b_{occ}) \quad (19)$$

where $\max(b_{free})=b_{snd}/3$ and $\max(b_{occ})=(1+\beta)CW$, we have
$$b_{snd} \geq b_{snd}/3+(1+\beta)CW \quad (20)$$
After rearrangement we have
$$b_{snd} \geq 3(1+\beta)CW/2 \quad (21)$$
where $\beta=1$ maximally, e.g., $\beta=1$ for QARR/BARR as it does not decrease *ssthresh*. Substitute $\beta=1$ into (21) we have
$$b_{snd} \geq 3CW \quad (22)$$
which shows that $b_{snd}$ must be at least 3 times of the current CWnd size to make sure that application stall will not happen during the loss recovery phase. ∎

*D. Performance Validation*

With the refined sending buffer growth mechanism, we revalidate the system model in Section IV.A using the same experimental setup. As shown in Table 4 and Table 5, the numerical results computed from the system model become consistent with the experimental results after applying the proposed optimization algorithms, i.e., opportunistic retransmission and refined sending buffer growth mechanism, which validates the correctness of our model and the efficiency of the proposed optimization algorithms. Note that the small gap from the 100% utilization for QARR is due to scheduling delay in the both sender and receiver's kernel according to our preliminary investigation, as our model assumes zero processing delay, which warrants further research.

## VI. IMPLEMENTATION

In this section, we implement the optimization algorithms, i.e., opportunistic retransmission and refined *sndbuf* growth mechanism, in the current Linux kernel. To facilitate the comparative evaluation of the existing loss recovery algorithms w/ and w/o the proposed optimization algorithms and the future practical deployment, we modularized the loss recovery algorithms so that those loss recovery algorithms can be implemented and switched in-and-out easily without recompiling the kernel. Specifically, we implemented all loss recovery algorithms, i.e., RH, PRR, QARR and BARR, in a separate kernel module respectively. The module handler interfaces are defined in *struct tcp_retrans_ops*, which has the similar implementation as the pluggable TCP congestion control module handler interfaces [29]. The *struct tcp_retrans_ops* is defined as follows:

```
struct tcp_retrans_ops {
  struct list_head list;
  /*initialize private data when entry CWR or Recovery mode (required)*/
  void (*init_cwnd_reduction)(struct sock *sk, const bool set_ssthresh);
  /* change cwnd based on packets newly delivered (required) */
  void (*cwnd_reduction)(struct sock *sl, int newly_acked_sacked,
                         int fast_rexmit);
  /* cleanup when exiting CWR or Recovery mode (required) */
  void (*end_cwnd_reduction)(struct sock *sk);
  /* hooker for packet ack accounting (optional) */
  void (*pkts_acked)(struct sock *sk, u32 num_acked, s32 rtt_us);
  /*opportunistic retransmission is handled here*/
  u32 (*get_awnd_extension)(const struct tcp_soack *tp);
  char name[TCP_RETRANS_NAME_MAX];
  struct module *owner;
};
```

In this *struct*, every loss recovery algorithm mainly requires two components: (1) the initialization of TCP parameters, e.g., *ssthresh*, when entering/exiting from the loss recovery phase, which are defined in the handlers of *init_cwnd_reduction* and *end_cwnd_reduction*; (2) the handler for each received ACK/DUPACK during the loss recovery phase, which is defined in *cwnd_reduction*. The loss recovery algorithm can enable/disable opportunistic retransmission by defining handler *get_awnd_extension*. We also introduce the *proc* interfaces used for *sndbuf* management and performance tunings: (a) *snd_buffer_grow_factor*: the *sndbuf* increasing factor, 3 by default; (b) *queue_len_threshold*: the target queue length in QARR, 5 by default (i.e., $Q_T=5$); (c) *slide_wnd_size*: the sliding window size for bandwidth estimations in BARR, 200 by default (i.e., $M=200$).

## VII. PERFORMANCE EVALUATIONS

In this section, we evaluate the performance impact of the proposed optimization algorithms on the overall TCP throughput in the presence of wide network settings, instead of only focusing on the throughput performance over the TCP loss recovery phase under the loss event of consecutive packet losses.

*A. Network Testbed Setup*

We setup a testbed as depicted in Fig. 2. We employ an emulator rather than conducting experiments over a production of mobile networks as it is not feasible to control packet loss events, therefore making consistent performance comparisons difficult. To emulate various network behaviors in terms of the bandwidth, loss rate and delay, we develop a custom emulator using DPDK [30], a set of libraries and drivers for fast packet processing, thus confining the outgoing rate at the short time scale without introducing packet bursts greater than the expected one.

In configuring the emulator we adopt typical network parameters including delay, loss, and bandwidth according to the recent measurements in [3, 12, 13, 31, 32], which are summarized in Table 3 if not mentioned. Specifically, to model loss events, the emulator *randomly* drops the outgoing packets. Three loss rates, i.e., 0.01%, 0.05% and 0.1% are tested in our experiments. We believe that these loss rates can cover the ones measured/estimated from production networks, which are reported in one recent measurement study [31], depending on the network configurations and radio signal conditions, including some extreme cases such as severe signal quality degradation and hands-off to another cell. To model the bandwidth, two types of mobile data networks are emulated: 3G/HSPA+ and 4G/LTE, with their capacities are set to 20Mbps and 100Mbps respectively according to [3, 31]. The base RTT estimated from productions networks is 150~200ms [12, 13, 32] and 60ms [31] in 3G and 4G respectively, thus four propagation delays, i.e., 50ms, 100ms 150ms and 200ms, are tested, which cover the ones measured.

The sender host, emulation host and receiver host in Fig. 2 run Centos 5.5 Linux with kernel v.3.10 with dual Xeon E5645 2.40GHz CPU, 32GB RAM and 1Gbps NICs. The peak sending throughput can reach 950 Mbps. We add the loss recovery modules into the sender's kernel, which can switch between different loss recovery algorithms w/ and w/o the proposed optimization algorithms.

## B. Module Verification

With the testbed, we can capture the TCP data trace and the TCP congestion control related parameters at the TCP sender, e.g., logging CWnd and *ssthresh* with *tcpprobe* module. In this section, we verify the correctness of the pluggable TCP loss recovery module. First, we compare the throughput results using the loss recovery module with the ones using the built-in implementation. We first evaluate the average throughputs of a single TCP flow under different loss rates, ranging from 0.001% to 10%. The congestion control and loss recovery algorithms are TCP CUBIC and PRR respectively. As shown in Fig. 3 (left), the average TCP throughputs under all the loss rates with the pluggable PRR module match the ones with the Linux built-in PRR. Second, we intentionally drop packets at five specific timings during the transmission of a TCP flow with and without using pluggable PRR module respectively. As shown in Fig. 3 (right), both CWnd dynamics also match with each other. The other loss recovery modules (i.e., RH, QARR and BARR) are also shown to have the consistent observation with the same verification method. In summary, the pluggable TCP loss recovery module offers an accurate tool to design, develop and test various TCP loss recovery algorithms.

## C. Impact of Loss Recovery Algorithms on TCP Throughput Performances

The results so far reveal that the existing TCP loss recovery algorithms are not able to efficiently utilize the bandwidth during loss recovery phase, e.g., 0.028 to 0.114 bandwidth utilization for QARR. However, during the period of no packet loss, TCP could achieve much higher bandwidth utilization, e.g., 19.3 Mbps, for no-loss case, thus achieving 96.5% bandwidth utilization over emulated 3G/HSPA+ network. We note that the 3.5% (i.e., 100%–96.5%=3.5%) bandwidth degradation is primarily due to the TCP congestion control bottleneck, e.g., the slow start phase. Intuitively TCP throughput degradation will increase as packet loss rate increases, as the TCP flow will spend more time operating in the loss recovery phase.

In this section, we first evaluate the baseline long-term TCP performance using 16 combinations of TCP variants, i.e., TCP CUBIC, TCP Westwood, TCP Veno and TCP Vegas, and loss recovery algorithms, i.e., RH, PRR, QARR and BARR, for 200 s under different loss rates, over emulated 3G/HSPA+ network. The reason why we choose in this way is that TCP CUBIC is the default TCP variant in Linux kernel implementation, and TCP Westwood and TCP Veno are designed for mobile/wireless network to combat random loss, and TCP Vegas is a representative of the delay-based TCP. RH is the default loss recovery algorithm before Linux Kernel 3.2, while PRR is the default loss recovery after Linux Kernel 3.2. Both QARR and BARR are widely employed in recent proposed rate control algorithms [3-11].

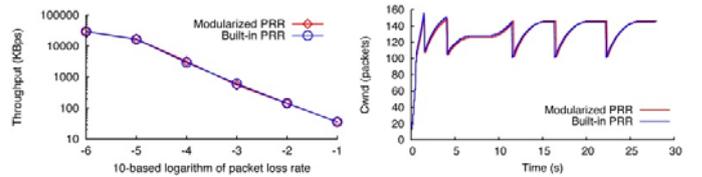

Fig. 3 – Pluggable TCP Loss recovery kernel module verification.

Table 6 – Throughput performance of the existing combinations of congestion control and loss recovery algorithms under different packet loss rates

| Congestion Control | Loss recovery | Loss rate (%) | | |
|---|---|---|---|---|
| | | 0.01 | 0.05 | 0.1 |
| CUBIC | RH | 17.0(85.1%) | 6.87(34.3%) | 5.2 (26.0%) |
| | PRR | 16.1(80.3%) | 7.89(39.5%) | 5.4 (27.1%) |
| | QARR | 18.8(94.0%) | 17.6(88.0%) | 15.6(78.5%) |
| | BARR | 18.7(93.5%) | 17.9(89.5%) | 16.6(83.0%) |
| Veno | RH | 14.2(71.0%) | 6.35(31.8%) | 4.72(23.6%) |
| | PRR | 14.1(70.6%) | 7.20(36.0%) | 5.3 (26.4%) |
| | QARR | 18.9(94.5%) | 16.6(83.0%) | 14.1(70.5%) |
| | BARR | 18.8(94.0%) | 17.9(89.5%) | 16.5(82.5%) |
| Westwood | RH | 14.6(72.5%) | 12.2(61.2%) | 7.87(39.4%) |
| | PRR | 16.7(83.5%) | 9.57(47.9%) | 8.30(41.6%) |
| | QARR | 18.6(93.0%) | 16.8(84.0%) | 15.3(77.0%) |
| | BARR | 18.2(91.0%) | 16.8(84.0%) | 16.2(81.0%) |
| Vegas | RH | 13.0(65.0%) | 6.69(33.5%) | 4.62(23.1%) |
| | PRR | 13.6(68.9%) | 6.07(30.4%) | 4.36(21.8%) |
| | QARR | 18.3(91.6%) | 16.9(84.5%) | 14.3(71.5%) |
| | BARR | 18.7(93.5%) | 17.6(88.0%) | 16.2(81.0%) |

As shown in Table 6, comparing to the no-loss case, TCP throughput degrades as the packet loss rate is increased. Even at a loss rate of 0.1%, which is not uncommon in mobile data networks, both TCP CUBIC+RH and TCP CUBIC+PRR's throughput drops from 17.0Mbps to 5.2Mbps, because TCP CUBIC blindly cuts down *ssthresh* and CWnd by 0.3, thus reduces the throughput. Surprise, TCP Veno, and TCP Westwood, that are optimized for mobile/wireless network, still suffer from the throughput degradation due to random loss, e.g., TCP Veno+RH/PRR and TCP Westwood+RH/PRR's throughputs dropped from 14.2Mbps to 5.3Mbps and 16.7Mbps to 7.87Mbps, respectively. After investigating the trace data we find out that they are not able to figure out an effective *ssthesh*/CWnd upon loss packets, leading to reducing CWnd unnecessarily. TCP Vegas+RH/PRR also reduces its CWnd/*ssthresh* blindly in a similar way. In summary, the throughputs of *all* TCP congestion controls with RH/PRR (TCP+RH/PRR) are throttled primarily by the congestion control bottleneck, e.g., CWnd/*ssthresh*, and their CWnds/*ssthresh* values are much less than AWnd reported, thus they do not suffer from the flow control bottleneck.

By contrast, all TCP+QARR/BARR combinations (e.g., TCP CUBIC+QARR/BARR) performed better than the other combinations under all loss rates but the throughput still degrades as the loss rate is increased (e.g., from 18.9Mbps to 14.1Mbps at 0.1% loss). It is worth noting that their CWnd/*ssthresh*s are unaffected by packet losses, while the throughput degradation in TCP+QARR/BARR in the presence of packet losses is primarily resulted from the low bandwidth utilization of the loss recovery phase due to the flow control bottleneck and application stall, rather than due to the congestion control.

Table 7 – Throughput performance of the existing combinations of congestion control and loss recovery algorithms under different propagation delays

| Congestion Control | Loss recovery | Propagation delay (ms) | | |
|---|---|---|---|---|
| | | 50 | 150 | 200 |
| CUBIC | RH | 6.98(34.9%) | 4.19(20.9%) | 3.76(18.8%) |
| | PRR | 7.25(36.2%) | 4.63(23.1%) | 3.93(19.6%) |
| | QARR | 18.0(90.0%) | 16.1(80.5%) | 14.7(73.5%) |
| | BARR | 18.0(90.0%) | 16.2(81.0%) | 14.9(74.5%) |
| Veno | RH | 9.58(47.9%) | 3.51(17.5%) | 2.52(12.6%) |
| | PRR | 10.3(51.5%) | 3.22(16.1%) | 2.38(11.9%) |
| | QARR | 18.3(91.5%) | 11.5(57.5%) | 10.1(50.5%) |
| | BARR | 18.1(90.5%) | 15.6(78.0%) | 14.5(72.5%) |
| Westwood | RH | 18.3(91.5%) | 7.64(38.2%) | 5.70(28.5%) |
| | PRR | 17.7(88.5%) | 6.86(34.3%) | 4.92(24.6%) |
| | QARR | 17.8(89.0%) | 14.0(70.0%) | 12.7(63.5%) |
| | BARR | 17.7(88.5%) | 14.4(72.0%) | 13.0(65.0%) |
| Vegas | RH | 9.60(48.0%) | 3.17(15.9%) | 2.28(11.4%) |
| | PRR | 9.85(49.2%) | 3.03(15.1%) | 2.31(11.5%) |
| | QARR | 17.3(86.5%) | 11.7(58.5%) | 11.4(57.0%) |
| | BARR | 17.7(88.5%) | 14.6(78.0%) | 14.5(72.5%) |

Next, we evaluate the baseline TCP throughput performances of those 16 combinations under different propagation delays. The analysis in Section III reveals that one of the primary sources of link idle time is the waiting time for the ACK for the first retransmitted packet to return to the sender. Thus, one would expect the propagation delay to have a significant impact on bandwidth efficiency during the loss recovery phase. The results in Table 7 show that, as the propagation delay increases from 100 ms to 200 ms, for example, TCP Vegas+QARR's throughput w/o the optimization algorithms drops from 17.3Mbps to 11.4Mbps. After investigating the data trace we found that, besides the flow control bottleneck and application stall, it also occasionally suffers from the congestion control bottleneck resulting from QARR algorithms. That is because the flow control bottleneck and application stall prevents QARR from transmitting new TCP packets, resulting in $Q_i$ draining below $Q_T$. QARR observes it and fills that queue up by increasing its CWnd, thus transmits packets in burst (c.f., Section VII.E). Those burst packets are then translated to the RTT spikes, which unnecessarily makes QARR decrease its CWnd, thus underutilize the bandwidth. Similarly, the flow control bottleneck and application stall also confuse the BARR algorithm by preventing it from transmitting new packets. Thus, BARR underestimates the bandwidth based on the insufficient ACK/DUPACKs received, which reduces its CWnd unnecessarily.

### D. Impact of the Optimization Algorithms on TCP Throughput Performances

We evaluate the impact of the proposed optimization algorithms on TCP's throughput with respect to two system parameters, namely packet loss rate and propagation delay over the same network. We first consider the packet loss rate in Table 8. With opportunistic retransmission and refined *sndbuf* growth mechanism, the throughput performances of all TCP+QARR/BARR combinations efficiently utilize the bandwidth at all loss rates, e.g., at least 94.5% bandwidth utilization. For example, TCP Vegas+QARR's throughput degrades insignificantly even at a loss rate of 0.1%. As compared to the original TCP Vegas+QARR, the proposed optimization algorithms improve the throughput by 33.6% at a loss rate of 0.1%, enabling TCP Vegas+QARR to achieve over 95.5% bandwidth utilization under all loss rates, and we expect that the improvement will be higher with the increase in loss rate, e.g., over 37% improvement at 0.2% loss rate, but the loss rate higher than 0.1% might not be common over current mobile data networks as measured. As the bandwidth utilization computed includes the TCP slow start period, thus the bandwidth utilization will be higher without slow start period. Note that the throughput performances of all TCP+RH/PRR combinations are not improved by the proposed optimization algorithms, as their throughputs are limited by the congestion control bottleneck, e.g., CWnd/*ssthresh*, during most of the experiments, instead of the flow control bottleneck and application stall.

Next, we consider the propagation delay in Table 9. As compared to Table 7, with the proposed optimization algorithms the throughput degradation for all the combinations of TCP+QARR/BARR are insignificantly affected, e.g., from 19.1 Mbps to 17.9 Mbps, as QARR/BARR has sufficient data to transmit during the loss recovery period. Hence, that throughput degradation (i.e., 19.1–17.9=1.2Mbps) is primarily due to the congestion control bottleneck. The optimization algorithms have little improvements on the combinations of TCP+RH/PRR, as RH/PRR reduces CWnd/*ssthresh* upon every loss event thus prevents its throughput from increasing.

### E. Eliminate Bursty Transmission

Besides throughput degradations flow control bottleneck and application stall also result in a side-effect – *bursty transmission*. Specifically, it prevents TCP sender from transmitting new data during the loss recovery phase, but those data will be transmitted in burst after loss recovery phase, causing RTT spikes, which is not desirable for delay-sensitive application such as video conferencing, running at the same link bottleneck.

To demonstrate this problem, we initiate a TCP flow that downloads 50MB data and intentionally drop a packet at 5 specific locations respectively using iptables [33] to emulate 0.01% loss rate. We record the CWnd and RTT dynamics using the existing loss recovery algorithms. As shown in Fig. 4 (left), we observed that, after the loss recovery phase, RTT increases by over 50%, e.g., 150ms versus 100ms, using PRR. In contrast, those RTT spikes are mitigated completely with the proposed optimization algorithms, e.g., remaining at 100 ms as shown in Fig. 4 (right), as the new data transmissions are spaced out over the loss recovery period by resolving the flow control bottleneck and application stall.

This problem also exists in QARR and BARR. For example, the burst size in QARR is more than the one in PRR, as it is not limited by the congestion control bottleneck, e.g., *ssthresh*, but the flow control bottleneck. As shown in Fig. 5 (left), the RTT increases by over 80%, e.g., 180 ms versus 100 ms. Similarly, the RTT spikes are completely mitigated when applying the proposed optimization algorithms (Fig. 5 (right)).

Table 8 – Performance impact of the proposed optimization algorithms on existing loss recovery algorithms under different packet loss rates.

| Congestion Control | Loss recovery | Loss rate (%) | | |
|---|---|---|---|---|
| | | 0.01 | 0.05 | 0.1 |
| CUBIC | RH | 17.0(85.1%) | 7.8 (38.8%) | 5.28(26.2%) |
| | PRR | 16.3(81.5%) | 9.0 (45.1%) | 5.99(29.9%) |
| | QARR | 19.1(95.5%) | 19.1(95.5%) | 19.1(95.5%) |
| | BARR | 19.1(95.5%) | 19.1(95.5%) | 19.0(95.0%) |
| Veno | RH | 15.1(75.4%) | 7.7 (38.4%) | 5.60(28.0%) |
| | PRR | 15.7(78.5%) | 8.4 (41.9%) | 6.25(31.3%) |
| | QARR | 19.1(95.5%) | 19.1(95.5%) | 19.1(95.5%) |
| | BARR | 19.1(95.5%) | 19.1(95.5%) | 19.0(95.5%) |
| Westwood | RH | 16.7(83.3%) | 12.9(64.5%) | 8.20(41.0%) |
| | PRR | 18.3(91.6%) | 13.4(66.9%) | 9.09(45.5%) |
| | QARR | 19.1(95.5%) | 19.1(95.5%) | 19.1(95.5%) |
| | BARR | 19.1(95.5%) | 19.1(95.5%) | 19.1(95.5%) |
| Vegas | RH | 13.0(65.0%) | 7.10(35.5%) | 4.79(23.9%) |
| | PRR | 14.3(71.3%) | 6.82(34.1%) | 4.82(24.1%) |
| | QARR | 19.1(95.5%) | 19.1(95.5%) | 19.1(95.5%) |
| | BARR | 19.1(95.5%) | 19.1(95.5%) | 18.9(94.5%) |

Table 9 – Performance impact of the proposed optimization algorithms on existing loss recovery algorithms under different propagation delays.

| Congestion Control | Loss recovery | Propagation delay (ms) | | |
|---|---|---|---|---|
| | | 50 | 150 | 200 |
| CUBIC | RH | 6.77(33.8%) | 4.05(20.2%) | 3.66(18.3%) |
| | PRR | 6.64(33.2%) | 4.10(20.5%) | 3.87(19.3%) |
| | QARR | 19.1(95.5%) | 19.0(95.0%) | 19.0(95.0%) |
| | BARR | 19.1(95.5%) | 18.6(93.0%) | 18.7(93.5%) |
| Veno | RH | 9.96(49.8%) | 3.40(17.0%) | 2.54(12.7%) |
| | PRR | 9.85(49.2%) | 3.69(18.4%) | 2.42(12.1%) |
| | QARR | 19.1(95.5%) | 19.0(95.0%) | 19.0(95.0%) |
| | BARR | 19.1(95.5%) | 18.8(94.0%) | 18.9(94.5%) |
| Westwood | RH | 18.6(93.0%) | 7.68(38.4%) | 5.57(27.8%) |
| | PRR | 17.7(88.5%) | 7.37(36.8%) | 5.45(27.2%) |
| | QARR | 19.1(95.5%) | 19.0(95.0%) | 19.0(95.0%) |
| | BARR | 19.1(95.5%) | 18.5(92.5%) | 17.9(89.5%) |
| Vegas | RH | 9.60(48.0%) | 3.17(15.9%) | 2.28(11.4%) |
| | PRR | 9.85(49.2%) | 3.03(15.1%) | 2.31(11.5%) |
| | QARR | 19.1(95.5%) | 19.1(95.5%) | 19.0(95.0%) |
| | BARR | 19.1(95.5%) | 18.7(93.5%) | 18.7(93.5%) |

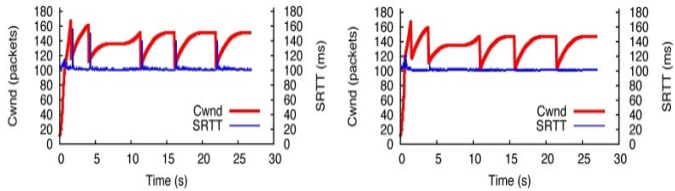

Fig. 4 – The CWnd and RTT results in downloading 50MB file with 5 packet loss. (left) CUBIC+PRR; (right) CUBIC+PRR w/ the proposed optimization algorithms.

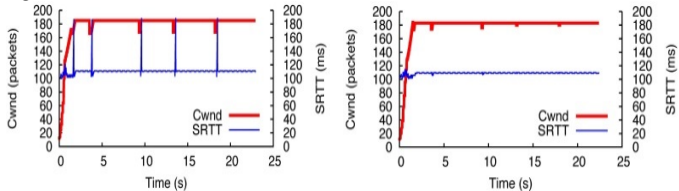

Fig. 5 – The CWnd and RTT results in downloading 50MB file with 5 packet loss. (left) CUBIC+QARR; (right) CUBIC+QARR w/ the proposed optimization algorithms.

### F. TCP Throughput Performances over 4G/LTE networks

The performance evaluations so far reveal that the optimized loss recovery algorithms efficiently utilize the bandwidth over lossy 3G/HSPA+ networks. It is also meaningful to evaluate it over high-speed mobile data networks, such as 4G/LTE networks, with the world-wide deployment of high-speed mobile data networks that offers over 100 Mbps link speed [3, 12, 31]. In this section, we evaluate the existing loss recovery algorithms w/ and w/o the optimization algorithms over 4G/LTE under the same setups except that the bandwidth of the emulator (i.e., *C* in Table 3) is set to 100 Mbps. The throughput results are summarized in Table 10-13. We omit the throughput results for TCP+PRR/RH combinations, because their throughputs are throttled primarily by the congestion control bottleneck.

Compared to the 3G/HSPA+, TCP first needs a longer ramp-up period to efficiently utilize the bandwidth over the 4G/LTE network. This period becomes even longer if the loss rate increases, as TCP spends more time in the loss recovery period over which its CWnd does not increase. We tested that a period of 200 s is long enough for TCP to reach a stable throughput. Second, the bandwidth loss due to the flow control bottleneck increases with the higher link bandwidth (c.f., Section III.F). Thus, as shown in Table 10 and Table 11, the baseline bandwidth efficiencies of all the TCP+QARR/BARR combinations over 4G/LTE are further reduced compared to Table 6 and Table 7, e.g., 66.3% or below, at 0.1% loss rate and 100ms delay. We note that QARR performs better than BARR, because BARR significantly underestimates the bandwidth due to few ACK/DUPACKs received during the loss recovery phase, thus reducing CWnd significantly.

As shown in Table 12 and 13, all the TCP+QARR/BARR combinations can achieve over 89.3% bandwidth efficiency with the proposed optimization algorithms under all the loss rates and propagation delays. Compared to 3G/HSPA+, the performance gain of the optimization algorithms in 4G/LTE becomes more significant, e.g., 43.3% versus 33.6% for TCP Vegas+QARR at 0.1% loss rate and 100ms propagation delay, and keeps increasing with higher BDP, which shows that the optimization algorithms can work in a wider network setting, i.e., the larger *U* and *C*. Similarly, the proposed optimization algorithms mitigate the RTT spikes in 4G/LTE.

### VIII. OPTIMIZATION LIMITATIONS

The assumption behind opportunistic retransmission is that the retransmitted packets can be received successfully (c.f., Section IV.A), otherwise, the receiving window cannot be freed to receive the TCP new packets transmitted opportunistically, so render the receiver to discard those *out-of-AWnd* packets. However, retransmitted packets are *rarely* lost again by examining the data trace of the previous experiments even at the loss rate of 0.1%. In this section, we investigate the impact on the throughputs of the optimized loss recovery algorithms when the assumption is *not* true.

Intuitively, high loss rates can make that assumption false. Thus, the loss rate is increased to 0.5%, which is uncommon in mobile data networks, and the experiments in Section VII.D are repeated. The throughput results with different propagation delays are summarized in Table 14. Except for CUBIC+QARR/BARR, we omit the results of the other

TCP+QARR/BARR combinations as they have the similar observations. Compared to the throughput results at 0.1% loss rate in Table 13, CUBIC+QARR/BARR cannot efficiently utilize the bandwidth, e.g., 46.8% versus 92.7% for $U$=150ms as shown in Table 14 (Optimized). Analysis of the trace data reveals the following three performance bottlenecks:

  a) The congestion control bottleneck: TCP spends more time during the loss recovery phase thus needs a longer ramp-up period at a larger loss rate.

  b) The flow control bottleneck: the receiving window cannot be freed up for receiving new packets transmitted opportunistically if the assumption is *not* true, i.e., the retransmitted packet can be dropped. These *out-of-AWnd* packets will be discarded by the receiver, which reduces the case *without* opportunistic retransmission thus utilize the bandwidth inefficiently.

  c) TCP timeouts: the receiver sends back DUPACKs without *new* SACKs when receiving those *out-of-AWnd* packets, but these DUPACKs cannot trigger the second *fast retransmit* of the retransmitted packet at the sender, as TCP flow control ensures that all packets are received within the receiving window thus it does not recognize DUPACKs without new SACKs. As a result, the sender has to wait until that retransmitted packet timeouts, which further deteriorates the throughputs of the optimized QARR/BARR algorithms, possibly making it worse than the throughputs of the original loss recovery algorithms, e.g., 46.8Mbps versus 47.4Mbps, as shown in Table 14 (Optimized and Original).

To further validate the bottlenecks (b) and (c), we configure the emulator not to drop the retransmitted packets and repeat the above experiments. As shown in Table 14 (Optimized (Optimal)), the bottlenecks (b) and (c) cannot be observed in the trace data thus the *optimal* bandwidth efficiency of the optimized QARR/BARR can reach over 93%.

Bottleneck (b) is inevitable unless making the assumption true in most cases, e.g., the network operator assigns high priority to the retransmitted packets in resource allocation and scheduling, but this requires modifications or reconfigurations of any of the mobile network equipment such as RNC or Node-B, which warrants further investigations. However, Bottleneck (c) can be resolved by modifying the sender's kernel to simply make it recognize the DUPACKs without new SACKs, thus we have the following hacks:

*Hack 1: in the procedure of processing incoming ACK/DUPACKs, it counts the received DUPACKs without new SACKs in TCP_CA_CWR state (i.e., loss recovery phase), if the count exceeds dupthresh, e.g., 3, it triggers fast retransmit of the latest unacknowledged TCP packet.*

As shown in Table 14 (Optimized (w/ Hack 1)), the optimized QARR/BARR with Hack 1 always perform better than the original ones – a direct result of the TCP timeout elimination, while the performance gap from the optimal ones is only due to bottleneck (b) and becomes larger with larger propagation delay, i.e., BDP. This is because, given the same loss rate, it is more likely for retransmitted packets to be lost if the *pipe* (i.e., packets inflight) becomes larger thus less throughput gain obtained from opportunistic retransmission.

Table 10 – Throughput performance of the combinations of congestion control and loss recovery algorithms under different loss rates over 4G/LTE.

| Congestion Control | Loss recovery | Loss rate (%) | | |
|---|---|---|---|---|
| | | 0.01 | 0.05 | 0.1 |
| CUBIC | QARR | 88.5(88.5%) | 73.6(73.6%) | 65.7(65.7%) |
| | BARR | 87.8(87.8%) | 63.8(63.8%) | 52.7(52.7%) |
| Veno | QARR | 88.7(88.7%) | 73.4(73.5%) | 66.3(66.3%) |
| | BARR | 87.4(87.4%) | 65.2(65.2%) | 53.7(53.7%) |
| Westwood | QARR | 88.5(88.5%) | 67.1(67.1%) | 60.5(60.5%) |
| | BARR | 84.9(84.9%) | 62.8(62.8%) | 54.9(54.9%) |
| Vegas | QARR | 88.3(88.3%) | 73.3(73.3%) | 65.8(65.8%) |
| | BARR | 87.9(87.9%) | 64.3(64.3%) | 51.2(51.2%) |

Table 11 – Throughput performance of the combinations of congestion control and loss recovery algorithms under different propagation delays over 4G/LTE.

| Congestion Control | Loss Recovery | Propagation delay (ms) | | |
|---|---|---|---|---|
| | | 50 | 150 | 200 |
| CUBIC | QARR | 73.5(73.5%) | 58.5(58.5%) | 54.9(54.9%) |
| | BARR | 66.3(66.3%) | 45.0(45.0%) | 39.2(39.2%) |
| Veno | QARR | 73.9(73.9%) | 60.6(60.6%) | 57.0(57.0%) |
| | BARR | 67.0(67.0%) | 47.3(47.3%) | 41.3(41.3%) |
| Westwood | QARR | 67.7(67.7%) | 54.9(54.9%) | 48.7(48.7%) |
| | BARR | 65.1(65.1%) | 43.8(43.8%) | 41.4(41.4%) |
| Vegas | QARR | 73.1(73.1%) | 57.3(57.3%) | 51.0(51.0%) |
| | BARR | 66.0(66.0%) | 43.8(43.8%) | 39.5(39.5%) |

Table 12 – Performance impact of the optimization algorithms on existing loss recovery algorithms under different packet loss rates over 4G/LTE.

| Congestion Control | Loss Recovery | Loss rate (%) | | |
|---|---|---|---|---|
| | | 0.01 | 0.05 | 0.1 |
| CUBIC | QARR | 95.2(95.2%) | 94.3(94.3%) | 93.9(93.9%) |
| | BARR | 95.2(95.2%) | 93.9(93.9%) | 93.1(93.1%) |
| Veno | QARR | 95.1(95.1%) | 95.1(95.1%) | 94.9(94.9%) |
| | BARR | 95.3(95.3%) | 95.1(95.1%) | 94.5(94.5%) |
| Westwood | QARR | 95.3(95.3%) | 93.3(93.3%) | 92.5(92.5%) |
| | BARR | 93.3(93.3%) | 92.8(92.8%) | 90.4(90.4%) |
| Vegas | QARR | 95.3(95.3%) | 95.0(95.0%) | 94.3(94.3%) |
| | BARR | 95.2(95.2%) | 95.0(95.0%) | 94.6(94.6%) |

Table 13 – Performance impact of the optimization algorithms on existing loss recovery algorithms under different propagation delays over 4G/LTE.

| Congestion Control | Loss Recovery | Propagation delay (ms) | | |
|---|---|---|---|---|
| | | 50 | 150 | 200 |
| CUBIC | QARR | 95.3(95.3%) | 92.7(92.7%) | 91.5(91.5%) |
| | BARR | 94.8(94.8%) | 92.1(92.1%) | 89.6(89.6%) |
| Veno | QARR | 95.3(95.3%) | 94.1(94.1%) | 93.3(93.3%) |
| | BARR | 95.3(95.3%) | 93.8(93.8%) | 92.2(92.2%) |
| Westwood | QARR | 95.1(95.1%) | 92.5(92.5%) | 89.3(89.3%) |
| | BARR | 94.4(94.4%) | 90.8(90.8%) | 90.7(90.7%) |
| Vegas | QARR | 95.3(95.3%) | 93.8(93.8%) | 91.4(91.4%) |
| | BARR | 95.4(95.4%) | 94.1(94.1%) | 92.1(92.1%) |

Table 14 – Performance impact of the optimization algorithms on QARR/BARR under different propagation delays at 0.5% loss rate.

| | Loss Recovery | Propagation delay (ms) | | | |
|---|---|---|---|---|---|
| | | 50 | 100 | 150 | 200 |
| Optimized | QARR | 84.3(84%) | 52.7(53%) | 46.8(47%) | 46.2(46%) |
| | BARR | 82.7(83%) | 51.6(52%) | 35.8(36%) | 33.7(34%) |
| Original | QARR | 54.5(55%) | 51.0(51%) | 47.4(47%) | 41.1(41%) |
| | BARR | 47.8(48%) | 34.9(35%) | 29.1(29%) | 26.2(26%) |
| Optimized (Optimal) | QARR | 95.2(95%) | 95.0(95%) | 94.4(94%) | 93.9(94%) |
| | BARR | 95.0(95%) | 94.4(94%) | 93.9(94%) | 93.5(94%) |
| Optimized (w/ Hack 1) | QARR | 85.3(85%) | 80.6(81%) | 76.0(76%) | 72.9(73%) |
| | BARR | 85.7(86%) | 79.0(79%) | 74.1(74%) | 71.8(72%) |

Table 15 – Performance impact of the optimization algorithms on QARR/BARR under various BDPs at 0.1% loss rate.

| Loss Recovery | | BDPs | | | |
|---|---|---|---|---|---|
| | | 100ms 200Mbps | 100ms 400Mbps | 100ms 800Mbps | 100ms 1Gbps |
| Original | QARR | 115(58%) | 199(50%) | 434(54%) | 502(50%) |
| | BARR | 98.9(49%) | 155(39%) | 224(28%) | 277(28%) |
| Optimized (Optimal) | QARR | 190(95%) | 371(93%) | 702(88%) | 929(93%) |
| | BARR | 188(94%) | 372(93%) | 713(89%) | 865(87%) |
| Optimized (w/ Hack 1) | QARR | 184 (92%) | 345(86%) | 658(81%) | 774(77%) |
| | BARR | 182 (91%) | 329(83%) | 642(80%) | 746(75%) |

On the other hand, besides the loss rate, the assumption also can be *false*, i.e., bottleneck (b) occurs, if the BDP is large enough, even at a lower loss rate. To validate this, we further increase the bandwidth and repeat the experiments at 0.1% loss rate. As shown in Table 15, the bandwidth efficiencies of the optimized QARR/BARR decreases with the increase in BDP, while the optimal bandwidth efficiencies of the optimized QARR/BARR without dropping retransmitted packets always remain at over 92%. Therefore, we conjecture that, for any loss rate, a *minimum* BDP can be derived theoretically that a predefined minimum bandwidth efficiency cannot be guaranteed (e.g., 80%) with the optimized TCP+QARR/BARR combinations, which warrants further investigations.

## IX.  SUMMARY AND FUTURE WORKS

In this work we proposed two optimization algorithms: opportunistic retransmission and refined buffer growth mechanism, to tackle the flow control bottleneck and application stall. The proposed algorithms significantly improve the throughput performances of the existing loss recovery algorithms and mitigate the RTT spikes that would occur after loss recovery phase. We are currently extending this work in two directions. First, we are extending the analytical model in Section III to characterize TCP's overall bandwidth utilization under various loss rates and BDPs. Second, we are applying the optimization algorithms for other networks which have larger BDPs and lower loss rates, such as datacenter, optical network and inter-ISP WAN.

## REFERENCES

[1]  S. Mascolo, C. Casetti, M. Geria, M. Y. Sanadidi and R. Wang, "TCP Westwood: Bandwidth Estimation for Enhanced Transport over Wireless Links," in *Proc. ACM MobiCom*, pp. 287-297, Jul 16 2001.

[2]  C. Fu, S. Liew, "TCP Veno: TCP Enhancement for Transmission over Wireless Access Networks," *IEEE JSAC*, vol. 21(2), pp.216-228, March 2003.

[3]  K. Liu and J. Y. B. Lee, "On Improving TCP Performance in Mobile Data Networks," IEEE Transactions on Mobile Computing, vol. 15, no. 10, pp. 2522-2536, Oct 2016.

[4]  F. Ren and C. Lin, "Modeling and Improving TCP Performance over Cellular Link with Variable Bandwidth," in *IEEE TMC*, vol. 10(8), pp.1057-1070, Aug 2011.

[5]  R. Chakravorty, S. Katti, I. Pratt and J. Crowcroft, "Using TCP Flow-Aggregation to Enhance Data Experience of Cellular Wireless Users," in *IEEE JSAC*, vol. 23(6), pp.1190-1204, Jun 2005.

[6]  M. Ivanovich, P. W. Bickerdike, and J. C. Li, "On TCP Performance Enhancing Proxies in a Wireless Environment," *IEEE Comm. Magazine*, vol. 46(9), pp.76-83, Sep 2008.

[7]  W. K. Leong, Y. Xu, B. Leong, and Z. Wang, "Mitigating egregious ACK delays in cellular data networks by eliminating TCP ACK clocking," in *Proc. IEEE ICNP*, pp. 1-10, Oct 7-10 2013.

[8]  Ke Liu and Jack Y. B. lee, "Achieving high throughput and low delay by accurately regulating link queue length over mobile data network," in *Proc. IEEE WiMob*, pp. 562-569, Oct 8-10 2014.

[9]  E. H. Wu and M. Chen, "JTCP: Jitter-based TCP for Heterogeneous Wireless Networks," *IEEE JSAC*, vol. 22, no. 4, pp.757-766, May 2004.

[10]  K. Xu, Y. Tian and N. Ansari, "TCP-Jersey for Wireless IP Communications," *IEEE JSAC*, vol. 22(4), pp.747-756, May 2004.

[11]  K. Winstein, A. Sivaraman and H. Balakrishnan, "Stochastic Forecasts Achieve High Throughput and Low Delay over Cellular Networks," in *Proc. USENIX NSDI*, pp.459-471, Apr 2-5, 2013.

[12]  J. Huang, F. Qian, Y. Guo, Y. Zhou, Q. Xu, Z. M. Mao, S. Sen and O. Spatscheck, "An In-Depth Study of LTE: Effect of Network Protocol and Application Behavior on Performance," in *ACM SIGCOMM*, vol. 43, no. 4, pp. 363-374, Oct 2013.

[13]  H. Jiang, Y. Wang, K. Lee and I. Rhee, "DRWA: A Receiver-Centric Solution to Bufferbloat in Cellular Networks," in *IEEE TMC*, vol. 15, No. 11, pp. 2719-2734, Nov 2016.

[14]  Y. Xu, W. K. Leong, B. Leong and A. Razeen, "Dynamic Regulation of Mobile 3G/HSPA Uplink Buffer with Receiver-Side Flow Control," in *Proc. IEEE ICNP*, pp.1-10, Oct 30 2012.

[15]  N. Dukkipati, M. Mathis, Y. Cheng, and M. Ghobadi, "Proportional Rate Reduction for TCP," in *Proc. ACM IMC*, pp.155-170, Nov 2 2011.

[16]  F. Lu, H. Du, A. Jain, G. M. Voelker, A. C. Snoeren, and A. Terzis, "CQIC: Revisiting Cross-Layer Congestion Control for Cellular Networks," in *ACM HotMobile*, pp.45-50, Feb 12 2015.

[17]  B. Frank, I. Poese, Y. Lin, G. Smaragdakis, A. Feldmann, B. Maggs, J. Rake, S. Uhlig, and R. Weber, "Pushing CDN-ISP Collaboration to the Limit," in *ACM SIGCOMM CCR*, vol. 43, no. 3, pp.34-44, July 2013.

[18]  Y. Zaki, T. Potsch, J. Chen, L. Subramanian and G. Gorg, "Adaptive Congestion Control for Unpredictable Cellular Networks," in *ACM SIGCOMM*, vol. 45, no. 4, pp. 509-522, Oct 2015.

[19]  S. G. Hazel, J. Iyengar, and M. Kuehlewind, "Low Extra Delay Background Transport (LEDBAT)," in *RFC 6817*, December 2012.

[20]  J. Wang, J. Wen, J. Zhang and Y. Han, "TCP-FIT: An Improved TCP Congestion Control Algorithm and its Performance," in *Proc. IEEE INFOCOM*, pp. 2894-2902, April 10-15 2011.

[21]  E. Blanton, M. Allman, K. Fall and L. Wang, "A Conservative Selective Acknowledgement (SACK)-based Loss Recovery Algorithm for TCP," in *RFC 3517,* April 2003.

[22]  M. Mathis and J. Mahdavi, "TCP Rate-halving with Bounding Parameters," Available: http://www.psc.edu/networking/papers/FACKnotes/current/.

[23]  M. Mathis, J. Mahdavi, S. Floyd, and A. Romanow, "TCP Selective Acknowledgment Options," in *RFC 2018*, Oct. 1996.

[24]  M. Mathis, J. Mahdavi, "Refining TCP Congestion Control," in *ACM SIGCOMM Computer Communication Review*, vol. 26, no 4, pp. 281-291, Oct 1996.

[25]  S. Seth, M. A. Venkatesulu, "TCP/IP Architecture, Design and Implementation in Linux," Wiley-IEEE Computer Society Press, 2008.

[26]  K. Liu and J. Lee, "Mobile accelerator: A New Approach to Improve TCP Performance in Mobile Data Networks," in *Proc. IWCMC*, pp. 2174-2180, 4-8 July 2011.

[27]  (2013) The iPhone 5s Review: A7 SoC Explained. [Online]. Available: http://www.anandtech.com/show/7335/the-iphone-5s-review/2

[28]  (2014) CPU Usage Limiter for Linux. [Online]. Available: http://cpulimit.sourceforge.net

[29]  S. Arianfar, "TCP's Congestion Control Implementation in Linux Kernel," in *Proc. Seminar on Network Protocols in Operating Systems*, 2012.

[30]  DPDK, Available: http://www.dpdk.org.

[31]  Y. Chen, Y, Lim, R. Gibbens, E. Nahum, R. Khalili and D. Towsley, "A measurement-based Study of MultiPath TCP Performance over Wireless Networks," in *Proc. ACM IMC*, pp. 455-468, Oct 23-25, 2013.

[32]  J. Huang, Q. Xu, B. Tiwana, Z. M. Mao, M. Zhang, and P. Bahl, "Anatomizing Application Performance Differences on Smartphones," in *ACM MobiSys*, pp.165-178, Jun 15-18, 2010.

[33]  The netfilter.org: iptables project homepage. [Online]. Available: http://www.netfilter.org/projects/iptables/index.html.

[34]  Z. Zha, K. Liu, B. Fu and M. Chen, "Optimizing TCP Loss Recovery Performance over Mobile Data Networks," in *IEEE SECON*, pp. 471-479, Jun 22-25 2015.